\documentclass[twocolumn,3p,times]{elsarticle}
  
\usepackage{mathptmx, courier, pifont}
\usepackage[scaled=0.92]{helvet}
\usepackage[T1]{fontenc}
\usepackage{textcomp}
\usepackage{amsfonts}

\usepackage{epsfig}
\usepackage{amssymb}
\usepackage{graphicx}
\usepackage{amssymb}
\usepackage{graphicx}
\usepackage{mathtools, cuted}
\biboptions{numbers,sort&compress}

\begin{document}

\begin{frontmatter}

\title{A new spectroscopic probe to search for magic numbers at high-excitation energies}
\author[a]{Cebo Ngwetsheni}
\author[a]{Jos\'e Nicol\'as Orce\corref{mycorrespondingauthor}}
\cortext[mycorrespondingauthor]{Corresponding author}
\ead{jnorce@uwc.ac.za}
\ead[url]{https://nuclear.uwc.ac.za}
\address[a]{Department of Physics \& Astronomy, University of the Western Cape, Bellville-7535, South Africa}

\date{\today}

\begin{abstract}

Empirical drops in ground-state nuclear polarizabilities indicate deviations from the effect of giant dipole resonances and 
may reveal the presence of shell effects in semi-magic nuclei with neutron magic numbers $N=50$, 82 and 126. 
Similar  drops of polarizability  in the quasi-continuum of nuclei with, or close to, magic numbers  $N=28$, 50 and 82, could reflect the continuing influence of shell closures 
up to the nucleon separation energy. 
These findings open a new avenue to investigating magic numbers at high-excitation energies and strongly support recent large-scale shell-model calculations 
in the quasi-continuum region, which describe the origin of the low-energy enhancement of the 
photon strength function as induced paramagnetism. The nuclear-structure dependence of the photon-strength function asserts the 
generalized Brink-Axel hypothesis as more universal than originally  expected.

\end{abstract}

\end{frontmatter}


The ability for a nucleus to be polarized is a fortiori driven by the dynamics of the isovector giant dipole resonance ({\small GDR}). 
That is, the inter-penetrating motion of proton and neutron fluids out of phase~\cite{migdal}, which 
results from the symmetry energy in the Bethe-Weizs\"acker semi-empirical  
mass formula~\cite{weiz,bethe}, $a_{sym}(\rho_{_N}-\rho_{_Z})^2/\rho_{_A}$, 
acting as a restoring force~\cite{migdal,steinwedel}. Respectively,  $\rho_{_N}$ and $\rho_{_Z}$  are the mass densities of the  neutron and  proton fluids,  and 
$\rho_{_A}$ the sum of the separate densities. 
The {\small GDR} represents most of the absorption and emission of $\gamma$-ray photons by a nucleus and was the first quantum collective 
excitation  ever discovered in mesoscopic systems~\cite{BK}. The idea of \emph{giant resonances} was soon borrowed by atomic, molecular and solid-state physics 
(see e.g.~\cite{gr_all} 
and references therein); the {\small GDR} motion is akin to the plasmons in graphene, which enables strong confinement 
of electromagnetic energy at subwavelength scales~\cite{graphene}.   

Using the collective variable $\rho_{_Z}$ as the potential energy of the liquid drop, Migdal calculated 
the  electric dipole polarizability, $\alpha_{_{E1}}$, for the  ground state of nuclei
to be directly proportional to the size of the nucleus~\cite{migdal}, 
\begin{equation}
\alpha_{_{E1}}=\frac{e^2R^2A}{40a_{sym}}=2.25\times 10^{-3} A^{5/3} ~\mbox{fm$^3$},
\label{eq:alpha}
\end{equation}
where $a_{sym}=23$ MeV is the symmetry energy parameter  and $R=1.2A^{1/3}$ fm the radius of the nucleus with $A=N+Z$. 
Alternatively, $\alpha_{_{E1}}$ is well described by microscopic 
mean-field approaches using the random-phase approximation ({\small RPA}) with various effective interactions~\cite{bohigas,plb1,plb2}, 
and can be determined empirically with the use of second-order perturbation theory, 
\begin{eqnarray}
\alpha_{_{E1}}&=&2e^2\sum_n \frac{\langle i\parallel\hat{E1}\parallel n\rangle \langle n\parallel\hat{E1}\parallel 
i\rangle}{E_{\gamma}} \nonumber \\  &=& \frac{\hbar c}{2\pi^2}\sigma_{_{-2}},  
\label{eq:polar} 
\end{eqnarray}
with  $|i\rangle$ being the vector of the ground state connecting high-lying $|n\rangle$  states in the {\small GDR} region via $E1$ virtual excitations, 
and $\sigma_{_{-2}}$ the $(-2)$ moment of the total 
photo-absorption cross section~\cite{levinger2,migdal3} defined as,
\begin{eqnarray}
\sigma_{_{-2}}:=\int_{0}^{E_{\gamma_{max}}}\frac{\sigma_{total}(E_{\gamma})}{E_{\gamma}^{^2}}~dE_{\gamma}, 
\label{eq:sigma-2}
\end{eqnarray}
where the total photonuclear-absorption cross section, $\sigma_{total}(E_{\gamma})$, generally includes 
all $\sigma(\gamma,n)$ and $\sigma(\gamma,p)$ contributions~\cite{atlas}. 

Naturally,  total $\sigma_{_{-2}}$ values should include both electric and magnetic polarizability contributions, 
\begin{eqnarray}
\sigma_{_{-2}}=\frac{2\pi^2}{\hbar c}(\alpha_{_{E1}}+\chi_{_{M1}}),
\label{eq:polarmag} 
\end{eqnarray}
where $\chi_{_{M1}}$ is the  static magnetic dipole polarizability and considers the sum of the paramagnetic $\chi^{para}_{_{M1}}$ and diamagnetic $\chi^{dia}_{_{M1}}$ 
susceptibilities  of nuclei~\cite{knupfer2}, 
\begin{eqnarray}
\chi_{_{M1}}&=&\chi^{para}_{_{M1}}+\chi^{dia}_{_{M1}} \nonumber \\  &=& 2\sum_n \frac{\langle i\parallel\hat{M1}\parallel n\rangle \langle n\parallel\hat{M1}\parallel i\rangle}{E_{\gamma}}
-\frac{Ze^2}{6mc^2}\langle r^2\rangle.
\label{eq:m1}
\end{eqnarray}

According to the  independent-particle shell model ({\small IPM}), diamagnetism is dominant for nuclei with $A>60$~\cite{knupfer}, 
but has a negligible effect in $\sigma_{_{-2}}$ values.  
Paramagnetism dominates in light nuclei with the rise of permanent magnetic dipole moments and 
can, in contrast, contribute substantially to  $\sigma_{_{-2}}$ values for nuclei with $A<20$~\cite{knupfer}.



Because of the 1/E$^2_{\gamma}$ energy weighting  in Eq.~\ref{eq:sigma-2}, $\sigma_{_{-2}}$ values are extremely sensitive measures -- unlike $\sigma_{total}$ -- of 
low-energy long-range correlations in the nuclear wave functions, 
which are common feature for all nucleon-nucleon potentials, and fundamental for shell-model ({\small SM}) calculations of heavy nuclei~\cite{Orce Nb} using 
low-momentum interactions~\cite{bogner}. 
Intermediate and short-range correlations to the nuclear wave functions
from above the {\small GDR} region (e.g., nucleon resonances at $E_{_\gamma}\gtrsim 140$ MeV) 
have a negligible effect on $\sigma_{_{-2}}$ values~\cite{levinger,320MeV,kerst_price,ahrens}. 

Below the neutron separation threshold, the pygmy dipole resonance ({\small PDR}) in neutron-rich nuclei~\cite{paar}  
-- the {\small PDR} is an electric dipole  resonance  arising from the oscillation of a symmetric proton-neutron core 
against the neutron skin -- 
may  add a $\simeq5\%$ contribution to $\sigma_{_{-2}}$ values~\cite{vNC}. To a 
lesser extent,  soft resonances 
such as the $M1$ scissors mode and spin-flip may also contribute. 
A potentially larger effect to $\sigma_{_{-2}}$ values may arise 
from the low-energy enhancement ({\small LEE}) of the radiative or photon strength function  $f(E_{\gamma})$  --  indicating the ability of nuclei to emit 
and absorb photons with energy $E_{\gamma}$ --  observed 
at $E_{\gamma}\lessapprox4$ MeV~\cite{guttormsen,mathis1,larsen}. 
This Letter shows how the {\small LEE} and  {\small GDR} cross-section contributions 
affect  $\sigma_{_{-2}}$ values  and may provide evidence for the 
continuing influence of shell effects at high-excitation energies. 
Relevant consequences arise from these findings; for instance, the possibility to identify new magic numbers. 

The physical origin of the {\small LEE} 
remains ambiguous and its observation seems to be generally associated with  weakly deformed nuclei. 
It has  been  observed in nearly-spherical nuclei in the $A\approx50$ and 90 mass regions starting at $E_{\gamma}\approx3-4$ MeV. 
For heavy nuclei,  it is only found in $^{105}$Cd~\cite{larsen_Cd}, $^{138,139}$La~\cite{kheswa} and 
$^{151,153}$Sm~\cite{simon}, where  the {\small LEE} starts  at a lower $E_{\gamma}\approx2$ MeV. 
These findings assume the validity of the Brink-Axel hypothesis 
-- stating that $f(E_{\gamma})$ is independent of the particular nuclear structure and only depends on E$_{\gamma}$~\cite{brink,axel} -- 
which has been confirmed experimentally~\cite{validityBrink,larsen_brink}. 
The reason for not being observed in other heavy nuclei -- studied with the same experimental method -- could relate to 
the unprecedented sensitivity achieved by Simon and co-workers in $^{151,153}$Sm using high-purity germanium ({\small HPGe}) detectors 
in connection with bismuth germanate ({\small BGO}) shields~\cite{simon}.  
Another relevant finding is that the {\small LEE} presents a dominant dipole radiation~\cite{larsen,larsen_brink}, 
but whether its nature is either electric or magnetic remains unresolved~\cite{jones}. 
The recent polarization asymmetry measurements of $\gamma$ rays in $^{56}$Fe using {\small GRETINA} tracking detectors yields inconclusive results, although rather suggests 
an admixture of electric and magnetic dipole radiation, with a small bias towards a magnetic character at $E_{\gamma}=1.5-2.0$ MeV~\cite{jones}.

Two competing scenarios are proposed theoretically to explain the {\small LEE} anomaly. 
On one hand, Litvinova and Belov propose that the {\small LEE} in $f(E_{\gamma})$ occurs because of $E1$ 
excitations from the hot-quasicontinuum to the continuum region~\cite{litvinova_E1}. On the other hand, {\small SM}  
calculations predict that the {\small LEE} has a predominant magnetic-dipole $M1$ character. 
In particular, Schwengner, Frauendorf and Larsen suggest that the {\small LEE} arises from active high-$j$ proton and neutron orbits
near the Fermi surface with magnetic moments adding up coherently~\cite{schwengner_M1}. 
This is a similar mechanism to the 
magnetic rotation~\cite{schwengner_MR} or two-phonon mixed-symmetry states
found in nearly-spherical nuclei at about 3 MeV~\cite{pietralla,fransen}.
In a complementary picture, Brown and Larsen suggest that the {\small LEE} arises because of the large $M1$ diagonal matrix elements 
of high-$\ell$ orbitals~\cite{brown}. 
Additionally, Sieja computed both $E1$ and $M1$ strengths in $^{44}$Sc on equal footing  from large-scale {\small SM} calculations and 
also supported the $M1$ character of the {\small LEE} in the $A\sim50$ region against $E1$ contributions~\cite{sieja,sieja2}. 
Recently, large-scale {\small SM} calculations of neutron-rich $^{70}$Ni~\cite{70ni} and many other nuclei~\cite{midtbo}, using various effective interactions, 
also support the $M1$ character for the {\small LEE}. 

Recently, 
In principle, the validation of these {\small SM} predictions in the quasi-continuum region may be arguable as, for instance, they are structure dependent;  
hence, posing a fundamental question about the validity of the Brink-Axel hypothesis~\cite{brink,axel}. 

A priori, the comparison of the  {\small LEE} and the {\small GDR} built on ground states is somewhat misleading as the former corresponds to $\gamma$-ray transitions between excited states 
in the quasi-continuum, whereas the latter involves transitions to the ground state. 
Nonetheless, the study of (p,$\gamma$) and (n,$\gamma$) reactions for light nuclei and fusion-evaporation 
reactions for heavy nuclei
have shown that {\small GDR}s  can also be built on excited states ({\small GDR$^{exc}$})~\cite{GDRenergy,GDRreview,schiller}. 
In fact,  {\small GDRs$^{exc}$} 
present -- at least for moderate average temperature $T$ and  spin $J$ -- 
similar centroid energies, $E^{exc}_{_{GDR}}$, and resonance  strengths, $S^{exc}_{_{GDR}}$, 
relative to the Thomas-Reiche-Kuhn ({\small TRK}) $E1$ sum rule~\cite{levinger3}, 
as those found for the ground-state counterparts ({\small GDR$^{g.s.}$})~\cite{GDRenergy,GDRreview}. 
These similar features suggest a common physical origin for all {\small GDR}s in concordance with the Brink-Axel 
hypothesis, which also indicates that a {\small GDR} can be built on every state in a nucleus~\cite{brink,axel}. 
Moreover, the sum rules in Eqs.~\ref{eq:polar} and \ref{eq:m1} can 
also be applied to final excited states $|f\rangle$~\cite{hausser2,orce_10Be,12C_kumar}. 
Henceforth, we assume similar resonance strengths for {\small GDR}s built on the ground and excited states. 
This may  explain the nice fit between 
the high $\gamma$-ray energy part of the measured $f(E_{\gamma})$ and the left tail of the {\small GDR$^{g.s.}$} (see e.g.~\cite{larsen_V}).

In order to combine cross-section contributions from the {\small LEE} and {\small GDR} regions, 
we use the well-known relation~\cite{becvar}, 
\begin{equation}
 f(E_{\gamma}) = \frac{1}{g_{_J}\pi^2(\hbar c)^2} ~\frac{\sigma_{_{total}}(E_{\gamma})}{E_{\gamma}} ~\mbox{MeV$^{-3}$}, 
 \label{eq:converter}
\end{equation}
where $g_{_J} = \frac{2J_f+1}{2J_i +1}$ is the statistical factor, with $J_i$ and $J_f$  being the spins of the initial and final states, respectively.  
The  magnitude of $g_{_J}$ affects the estimation of $\sigma_{_{-2}}$ values in the {\small LEE} region. 
However, assuming a predominant dipole character for the {\small LEE} radiation~\cite{jones,larsen,larsen_brink},  a value $g_{_J}=1$ is valid for $J\rightarrow J$ dipole transitions 
and a good approximation for any $\Delta J=1$ spin distribution  typically 
populated (up to $J=8-10 \hbar$) in the experimental studies of $f(E_{\gamma})$~\cite{kheswa2}.
This approximation is not valid for {\small GDR} $E1$ transitions in 
even-even nuclei where $g_{_J}=3$ applies.

\begin{figure}[!ht]
\includegraphics[width=7.5cm,height=6cm,angle=-0]{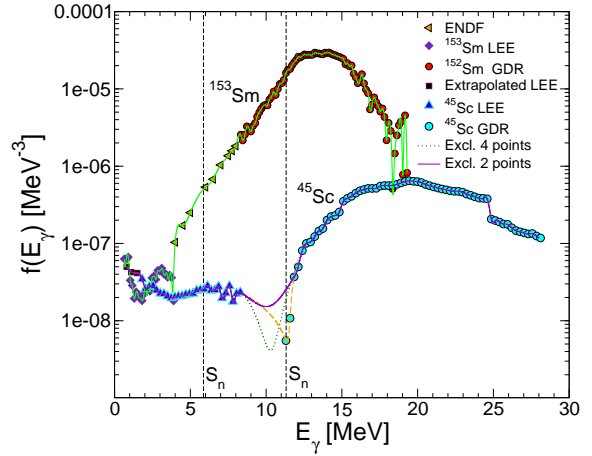}
\caption{$f(E_{\gamma})$ vs $E_{\gamma}$ on a log scale showing 
the interpolation  to the data (solid lines) for $^{45}$Sc~\cite{vessiere_Sc,larsen_Sc,shoda} and $^{153}$Sm~\cite{carlos_152Sm,simon}. 
Vertical dash lines indicate the neutron separation energy. See text for additional information.}
\label{comparison}
\end{figure}

The data spanning the {\small GDR} region have been obtained from available experimental nuclear reaction data bases, 
{\small EXFOR}~\cite{exfor} and {\small ENDF}~\cite{ensdf}. Data corresponding to the {\small LEE} -- in units of MeV$^{-3}$ -- 
have been collected from the Oslo compilation of level densities and $f(E_{\gamma})$~\cite{oslo}. 
The resulting $\sigma_{total}(E_{\gamma})$ was modeled using a cubic-spline interpolation -- which assumes validity of the Brink-Axel hypothesis -- 
in order to compute the total cross section and $\sigma_{_{-2}}$ values. 
Fourth-order polynomial fits yield similar results to the cubic spline interpolation, with almost negligible differences for the  integrated 
$\sigma_{_{-2}}$ values of <0.5\%. Lower and higher-order interpolation polynomials predict unanticipated structures of the ($\gamma$,n) cross-section 
(e.g. pronounced bumps between the {\small LEE} and {\small GDR} regions).

When available, {\small ENDF} data have been utilized to fill the typical gap between the {\small LEE} and {\small GDR} data sets, 
as shown in Fig.~\ref{comparison} for the case of $^{153}$Sm. Nuclei at different mass regions are evaluated 
for a systematic study of the {\small LEE} and {\small GDR}  effects on  $\sigma_{_{-2}}$ values. The results are listed in Table \ref{tab:LEE} and Fig.~\ref{comparison} shows the 
particular fits to the $^{45}$Sc  and  $^{153}$Sm  data. Uncertainties on $\sigma_{_{-2}}$ values 
arise from the {\small $RMS$} deviation,which accounts for a 7\% error from the lower and upper loci limits provided by {\small GDR} and {\small LEE} data~\cite{oslo}.

For the particular case of $^{45}$Sc, the large E$_{\gamma}$  gap between the {\small LEE} and {\small GDR} data 
resulted in unrealistic fits with a drastic drop of $\sigma_{_{-2}}$ values, as shown in Fig.~\ref{comparison}. Additional fits were performed by rejecting either 
the last two or the last four GDR data points at lower E$_{\gamma}$. Figure~\ref{comparison} shows that the former (solid line) is clearly more 
realistic and the resulting $\sigma_{_{-2}}$ value -- with a  4\% increase with respect to the fit considering the four {\small GDR} data points -- is quoted in Table~\ref{tab:LEE}. 
Fits to the data for the rest of nuclei studied in this work do not present such large energy gaps and re-fitting of the data was not found necessary.


\begin{table*}[!ht]
\centering
\begin{tabular}{l*{6}{c}r}
\hline \hline
Nucleus                & E$_{\gamma(max)}$(GDR)  &  E$_{\gamma(max)}$(LEE)        & $\sigma_{_{-2}}$(total)     & $\sigma_{_{-2}}$              & $\kappa$           &  [Refs.] \\
		       & (MeV)                   &  (MeV) 			  & ($\mu$b/MeV)                & (LEE)                         &  (LEE)             &             \\
\hline
 $^{45}_{21}$Sc$^{*}$  & 28.1                    & 	3.2  			  & 1840(130)                   & 9.7$\%$                       & 1.35(9)	      &  ~\cite{vessiere_Sc,larsen_Sc,shoda}  \\
 $^{50}_{23}$V         & 27.8                    &	3.1			  & 1458(100)	                & 2.9$\%$                       & 0.89(7)	      &  ~\cite{fultz_V,larsen_V}    \\
 $^{51}_{23}$V         & 27.8                    &	3.1			  & 1472(100)	                & 3.3$\%$                       & 0.87(7)	      &  ~\cite{fultz_V,larsen_V}    \\
 $^{56}_{26}$Fe$^{*}$  & 40.0                    &	3.8			  & 2231(155)                   & 6.3$\%$                       & 1.13(8)	      &  ~\cite{borodina_Fe,larsen}     \\
 $^{76}_{32}$Ge        & 26.5                    &      2.3 			  & 3189(225)                   & 2.7$\%$                       & 0.97(7)            &  ~\cite{Carlos_Ge,spyrou_Ge}  \\
 $^{92}_{40}$Zr        & 27.8                    &	2.2 			  & 3131(220)                   & 1.1$\%$                       & 0.70(5)            &  ~\cite{Berman_Zr,Guttormsen_Zr,utshumiya_Zr}  \\
 $^{95}_{42}$Mo        & 27.8                    &  	2.5  			  & 4743(330)                   & 1.7$\%$                       & 1.00(7)	      &  ~\cite{beil_Mo,utshumiya_Mo}     \\
 $^{138}_{\ 57}$La     & 24.3                    &      1.9 			  & 7983(560)                   & 0.4$\%$                       & 0.90(7)	      &  ~\cite{beil_La,kheswa}     \\
 $^{139}_{\ 57}$La     & 24.3                    & 	2.5			  & 8015(560)                   & 0.7$\%$                       & 0.90(6)	      &  ~\cite{beil_La,kheswa}      \\
 $^{153}_{\ 62}$Sm     & 20.0                    &	1.6 			  & 9999(700)                   & 2.7$\%$                       & 0.95(7)	      &  ~\cite{carlos_152Sm,simon}    \\
\hline \hline
\end{tabular} 
\caption{Contributions of {\small GDR} and {\small LEE} cross-sections to $\sigma_{_{-2}}$  and $\kappa$ values. 
Data have been extracted from  {\small EXFOR}~\cite{exfor}, {\small ENDF}~\cite{ensdf} and the Oslo compilation~\cite{oslo}. 
An asterisk indicates that the  $\sigma_{_{-2}}$ value includes  $\sigma(\gamma,p)$ contributions.}
\label{tab:LEE}
\end{table*}


Although the work by Jones and co-workers supports an increasing trend of {\small LEE} 
for $E_{\gamma}<1$ MeV~\cite{jones}, 
there is little  evidence on how  $f(E_{\gamma})$ behaves approaching $E_{\gamma}=0$. 
Hence, 
the low-energy cut off has arbitrarily been set to 800 keV for the nuclides considered in this work up to $^{139}$La, 
which incidentally is the typical energy for strong $M1$ isovector transitions 
in nearly-spherical nuclei~\cite{Orce Nb}.  
For $^{153}$Sm, a low-energy cut off of 645 keV has been set  from $f(E_{\gamma})$ data~\cite{simon}. 
Because of their instability, there is no available {\small GDR} information in $^{153}$Sm, $^{138}$La and $^{50}$V, 
and, instead, {\small GDR} data from  $^{152}$Sm, $^{139}$La and $^{51}$V, respectively, 
have been used in the analysis, under the assumption that nearby isotopes 
present equal $f(E_{\gamma})$ (see e.g.~\cite{guttormsen} and ~\cite{guttormsen2}). 
This assumption may not be  adequate 
given the rapid shape transition 
from weakly deformed in $^{150}$Sm to a well-deformed rotor in $^{154}$Sm, and the realization of 
shell closures in $^{139}$La ($N=82$) and $^{51}$V ($N=28$).

\begin{figure}[!ht]
\includegraphics[width=8cm,height=6.5cm,angle=-0]{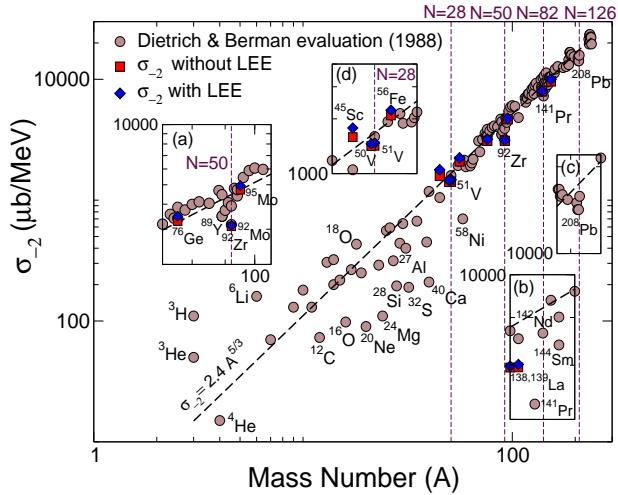}
\caption{$\sigma_{_{-2}}$ $vs$ A  on a log-log scale from the photo-neutron 
cross-section evaluation (solid circles)~\cite{atlas} and $\sigma_{-2}$ data listed in Table~\ref{tab:LEE} 
excluding (squares) and including (diamonds) the {\small LEE} contributions.  
For comparison, Eq.~\ref{eq:orce} (dashed line) is  plotted.}
\label{sigmaz}
\end{figure}

For comparison, Fig.~\ref{sigmaz} shows overall $\sigma_{_{-2}}$ values of ground states 
as a function of $A$ extracted from  photo-neutron cross sections  
using monoenergetic photon beams and determined above neutron threshold to an upper limit of $E_{\gamma_{max}}\approx20-50$ MeV~\cite{atlas}. 
The data include the {\small GDR} region and are representative for nuclei 
above $A\gtrapprox50$ (except for $^{58}$Ni~\cite{lectures}), where 
neutron emission is generally the predominant decay mode. 
This may not be true for nuclei with semi-magic number of neutrons --  discussed below --  where proton separation energies may lie lower 
than neutron thresholds.

%

From Eqs.~\ref{eq:alpha} and \ref{eq:polar}, Migdal extracted the relation $\sigma_{_{-2}}=2.25 A^{5/3} \mbox{$\mu$b/MeV}$, 
which was qualitatively confirmed by Levinger~\cite{levinger} and further refined~\cite{orce_p} as (dashed line in Fig.~\ref{sigmaz}), 
\begin{equation}
\sigma_{_{-2}}=2.4(1) \kappa~ A^{5/3} \mbox{$\mu$b/MeV}, 
\label{eq:orce}
\end{equation}
where $\kappa$ is the polarizability parameter and represents deviations from the actual {\small GDR} effects. 
This result is in excellent agreement with {\small IPM} predictions using, instead of $\hbar\omega=41A^{-1/3}$ MeV,   
$E^{g.s.}_{_{GDR}}= 79 A^{-1/3}$ MeV as the resonance frequency~\cite{bohigas2}. 
A value of $\kappa=1$ generally holds for the ground state of nuclei with $A\gtrapprox50$, and probably for even lighter nuclei with $A\gtrapprox20$ 
once  $\sigma(\gamma,p)$ contributions are taken into account~\cite{levinger,lectures,orce_p}. In contrast, values of $\kappa>1$ are generally found for 
light nuclei with $A<20$~\cite{levinger,orce_p,hausser2,10B_vermeer,barker,orce_10Be,12C_kumar}, where paramagnetism is important.

Sudden drops of $\sigma_{_{-2}}$ (and $\kappa$) values are apparent 
for the  $N=50$, 82 and 126 isotones 
in the insets (a), (b) and (c) of Fig.~\ref{sigmaz}, respectively. 
Above both proton and neutron separation energies, the photo-absorption cross section in the lower energy part of the {\small GDR} is controlled by 
the statistical competition between $\sigma(\gamma,p)$ and $\sigma(\gamma,n)$ contributions, which  presents a 
strong  correlation with the level density ratio $N_p/N_n$ between the open neutron ($N_n$) and proton ($N_p$) channels~\cite{lectures}, 
$\sigma(\gamma,p)/\sigma(\gamma,n)\approx N_p/N_n$.
This ratio depends on the neutron and proton penetrabilities, $\epsilon_n$ and $\epsilon_p$, respectively,  as more energy is needed for protons 
to overcome the Coulomb barrier. 

Total photo-absorption  cross-sections ($\sigma(\gamma,n)$ + $\sigma(\gamma,p)$) are  reasonably  available in the $N=50$ isotones, with the latter being indirectly determined 
from ($e,e^{\prime}p$) measurements~\cite{shoda}. 
The $\sigma(\gamma,p)$ contribution is particularly important for $^{92}$Mo, with $N_p/N_n\approx1.95$, 
and decreases for the lighter $N=50$ isotones, with $N_p/N_n\approx0.66, <0.28$ and 0.09 for $^{90}$Zr, $^{89}$Y and $^{88}$Sr, respectively~\cite{beil_Mo},  
as the isospin quantum number $T_z=\frac{N-Z}{2}$ increases. 
The  $\sigma(\gamma,p)$ contribution extracted from the $N_p/N_n$ ratio only applies to the lower energy half of the {\small GDR}, 
and $\sigma(\gamma,n)$ contributions  still remain greater. 
Once  $\sigma(\gamma,p)$ contributions are taken into account,  
the total photo-absorption cross section
satisfies the {\small TRK} sum rule~\cite{beil_Mo}, 
$\frac{\sigma_{total}(\gamma,n)+\sigma_{total}(\gamma,p)}{0.06NZA^{-1}}$. 
For the $^{92}$Mo case, there remains $\approx35\%$ $\sigma(\gamma,p)$ contribution to the total photo-absorption cross section~\cite{shoda}, 
which explains the sharper drop in the $\sigma_{_{-2}}$ value shown in Fig.~\ref{sigmaz}(a).
More conspicuous are the drops of $\sigma_{_{-2}}$ values in $^{89}$Y, $^{141}$Pr and $^{208}$Pb --  where 
$\sigma(\gamma,n)$ contributions  strongly dominate -- which could provide evidence for shell effects.
Clearly, direct measurements of $\sigma(\gamma,p)$ contributions are crucially needed for singly- and doubly-magic nuclei. 
Furthermore, Table~\ref{tab:LEE} shows that the {\small LEE} has a 
substantial contribution to $\sigma_{_{-2}}$ values in medium-mass nuclei ($^{45}$Sc and $^{56}$Fe) away from the 
$N=28$ shell closure, 
being largest for $^{45}$Sc with $\approx10\%$ increase. 
As illustrated in Fig.~\ref{comparison}, this  enhancement  partly arises 
because of the inverse mass dependence of $E_{_{GDR}}$ 
and the fact that the {\small LEE} starts at lower $E_{\gamma}$ as $A$ increases. 
In fact, Table~\ref{tab:LEE} shows that the {\small LEE} has a negligible contribution of $\lesssim3\%$  to the total $\sigma_{_{-2}}$ values of heavy nuclei with $A\geqq76$. 
A stronger contribution to $\sigma_{_{-2}}$  values would arise if the {\small LEE} trend keeps increasing  
at energies approaching $E_{\gamma}=0$, as predicted by {\small SM} calculations. This possibility will be explored in detail in a separate manuscript~\cite{cebo}.

More intriguing are the small overall contributions to $\sigma_{_{-2}}$ values found in nuclei close 
to or having a magic number. When compared with Eq.~\ref{eq:orce}, these nuclides present evident deviations from {\small GDR} effects (i.e. $\kappa\neq1$)
with smaller values of $\kappa\approx0.90$ in $^{50,51}$V ($N\approx28$) and $^{138,139}$La ($N\approx82$), 
and specially for $^{92}$Zr ($N\approx50$ and $Z=40$) with $\kappa=0.70(5)$. 
In contrast, heavy nuclei away from shell closures present polarizability parameters consistent with $\kappa=1$; 
except perhaps for $^{153}$Sm, where we used the $^{152}$Sm data for the {\small GDR} region and a cut-off of $E_{\gamma}=645$ keV. 
This recurrent behavior to the one previously observed in the photo-neutron cross-section data for the  $N=50$, 82 and 126 isotones, 
indicates the continuing influence of shell effects in the quasi-continuum region up to the neutron threshold.  
As shown in Table~\ref{tab:LEE} and inset (d) in Fig.~\ref{sigmaz}, this is consistent with 
the smaller {\small LEE} contribution to 
the total $\sigma_{_{-2}}$ values of $^{50,51}$V ($N\approx28$) with respect to the neighboring $^{45}$Sc and $^{56}$Fe nuclides.  
Although there is no $\sigma(\gamma,p)$ data available for $^{50,51}$V, $(\gamma,p)$ contributions 
will relatively be  much weaker for $^{51}$V because of the much lower level density 
of the  open proton channel (even-even $^{50}$Ti with $N=28$) as compared with the open neutron channel (odd-odd $^{50}$V). 

Interesting {\small SM} calculations of the $M1$ strength in the {\small LEE} for various isotopic and isotonic chains by Midtbo and collaborators~\cite{midtbo} 
predict a relatively sharper increase of the $M1$ strength at $E_{\gamma}=0-2$ MeV for neutron-rich nuclei when approaching shell closure. 
These results may contradict our findings for stable nuclei and suggest the enhancement of nuclear polarizability with increasing instability.

Conclusively, drops of $\sigma_{_{-2}}$ values ($\kappa<1$) for several nuclei with, or close to, neutron 
magic numbers $N=28$, 50, 82 and 126, suggest  that the shell model remains valid at high excitation energies, 
from the quasi-continuum to the {\small GDR} region; in agreement with Balashov's {\small SM} interpretation of the {\small GDR} 
as a system of independent nucleons plus the residual interaction~\cite{balashov}. 
These deviations from {\small GDR} effects, because of the nature of Eq.~\ref{eq:orce}, are plausibly not related to $E1$ transitions, 
which, together with the continuing influence of shell effects, strongly support  the $M1$ interpretation of the {\small LEE} by 
large-scale {\small SM} calculations~\cite{schwengner_M1,brown,sieja,sieja2,70ni}. 
Moreover, the empirical evidence for shell effects suggests that the generalized Brink-Axel hypothesis allows for 
structural changes and is, therefore, more universal than originally expected. 
This conclusion is supported by the work of Larsen {\it et al.}~\cite{larsen_brink}, where $f(E_{\gamma})$ trends 
are found to be preserved for different bin energies.

Finally, we confirm the induction of permanent magnetic dipole moments or paramagnetism in the quasi-continuum region, in agreement with previous 
{\small SM} calculations and {\small IPM}  predictions of 
an enhanced paramagnetism for the ground states of nuclei with large occupation number of the shells determining the 
magnetic properties~\cite{knupfer2}. The origin of this paramagnetism can be inferred from {\small SM} calculations, which  can distinguish 
between single-particle spin-flips and collective isovector excitations by decomposing the relevant $M1$ strength into their spin and orbital
components~\cite{Orce Nb}. 
Similar to two-neutron separation energies extracted from atomic mass measurements of ground and isomeric states, 
this work opens a new research avenue to investigate the evolution of shell closures and the existence of ``old'' and ``new'' magic numbers 
at high-excitation energies  from $\sigma_{_{-2}}$ measurements. 

The authors acknowledge Physics discussions with B. Dey, P. E. Garrett, B. V. Kheswa, S. Triambak, K. Sieja, P. von Neumann-Cosel and  M. Wiedeking. 
This work was supported by the  National Research 
Foundation of South Africa under Grant 93500 and the MaNuS/MatSci Honours/Masters program.



\begin{thebibliography}{100}


%



\bibitem{migdal} A. Migdal, J. Exptl. Theoret. Phys. U.S.S.R. {\bf 15} (1945) 81. 



\bibitem{weiz} C. F. von Weizs\"acker, Z. Phys. {\bf 96} (1935)  431. 

\bibitem{bethe} H. A. Bethe and R. F. Bacher, Rev. Mod. Phys. {\bf 8} (1936) 82.

\bibitem{steinwedel} H. Steinwedel, J. H. D. Jensen, and P. Jensen, Phys. Rev. {\bf 79} (1950) 1019.

\bibitem{BK} G. C. Baldwin and G. S. Klaiber, Phys. Rev. {\bf 71} (1947) 3.

\bibitem{gr_all} Giant resonances in atoms, molecules and solids, NATO ASI Series B, Physics {\bf 151} (1986). 


\bibitem{graphene}  M. Jablan, M. Solja\v{c}i\'c, and H. Buljan, Proc. of the IEEE {\bf 101} (2013) 1689. 




%
%
%
%
%
%


						  
						  

\bibitem{bohigas} O. Bohigas, N. van Giai, and D. Vautherin, Phys. Lett. B {\bf 102} (1981) 105.

\bibitem{plb1} Z. Zhang, Y. Lim, J. W. Holt, and C. M. Ko, Phys. Lett. B {\bf 777} (2018) 73.

\bibitem{plb2} D. Gambacurta, M. Grasso, and O. Vasseur, Phys. Lett. B {\bf 777} (2018) 163.




\bibitem{levinger2} J. S. Levinger, \emph{Nuclear Photo-Disintegration} (Oxford University Press, Oxford, 1960).



\bibitem{migdal3} A. B. Migdal, A. A. Lushnikov, and D. F. Zaretsky, Nucl. Phys. A {\bf 66} (1965) 193.


\bibitem{atlas} S. S. Dietrich and B. L. Berman, Atom. Data Nucl. Data Tables {\bf 38} (1988) 199. 


\bibitem{knupfer2} W. Kn\"upfer and A. Richter, Z. Phys. A {\bf 320} (1985) 253.


\bibitem{knupfer} W. Kn\"upfer and A. Richter, 	Phys. Lett. B {\bf 107} (1981) 325.	


\bibitem{Orce Nb} J. N. Orce {\it et al.}, Phys. Rev. Lett. {\bf 97} (2006)  062504.                                     

\bibitem{bogner} S. K. Bogner, R. J. Furnstahl, and A. Schwenk, Prog. in Part. Nucl. Phys. {\bf  65} (2010)  94. 


\bibitem{320MeV} L. W. Jones and K. M. Terwilliger, Phys. Rev.  {\bf 91} (1953) 699. 
\bibitem{kerst_price} D. W. Kerst and G. A. Price, Phys. Rev. {\bf  79} (1950)  725.
\bibitem{ahrens} J. Ahrens, H. Gimm, A. Zieger, and B. Ziegler, Il Nuovo Cimento A, Vol. {\bf 32} N. 3  (1976)  364.
\bibitem{levinger} J. S. Levinger, Phys. Rev. {\bf 107}  (1957) 554.


\bibitem{paar} N. Paar, D. Vretenar, E. Khan, and G. Colo, Rep. Prog. Phys. {\bf 70} (2007) 691. 

\bibitem{vNC}  P. von Neumann-Cosel, Phys. Rev. C {\bf 93} (2016) 049801. 



\bibitem{guttormsen} M. Guttormsen {\it et al.}, Phys. Rev. C {\bf 71} (2005) 044307. 
\bibitem{mathis1} M. Wiedeking {\it et al.}, Phys. Rev. Lett. {\bf 108} (2012) 162503.
\bibitem{larsen} A. C. Larsen {\it et al.}, Phys. Rev. Lett. {\bf 111} (2013) 242504. 




\bibitem{larsen_Cd} A. C. Larsen {\it et al.}, Phys. Rev. C {\bf 87} (2013) 014319.

\bibitem{kheswa} B. V. Kheswa {\it et al.}, Phys. Lett. B {\bf 744} (2015) 268.

\bibitem{simon} A. Simon {\it et al.,} Phys. Rev. C {\bf 93} (2016) 034303.   



\bibitem{brink} D. Brink, doctoral thesis, Oxford University, 1955 (unpublished). 

\bibitem{axel} P. Axel, Phys. Rev. {\bf 126} (1962) 671.                                                         


\bibitem{larsen_brink} A. C. Larsen {\it et al.}, J. Phys. G {\bf 44} (2017) 064005. 

\bibitem{validityBrink} M. Guttormsen {\it et al.}, Phys. Rev. Lett. {\bf 116} (2016) 012502.                     




\bibitem{jones} M. D. Jones {\it et al.}, Phys. Rev. C {\bf 97} (2018) 024327. 





\bibitem{litvinova_E1} E. Litvinova and N. Belov, Phys. Rev. C {\bf 88} (2013) 031302(R). 
\bibitem{schwengner_M1} R. Schwengner, S. Frauendorf and A. C. Larsen, Phys. Rev. Lett. {\bf 111} (2013) 232504. 
\bibitem{schwengner_MR} R. Schwengner {\it et al.}, Phys. Rev. C {\bf 66}  (2002) 024310.





\bibitem{pietralla} N. Pietralla {\it et al.}, Phys. Rev. Lett. {\bf 83} (1999) 1303.

\bibitem{fransen} C. Fransen {\it et al.}, Phys. Rev. C {\bf 67} (2003) 024307.                                  




\bibitem{brown} B. A. Brown and A. C. Larsen, Phys. Rev. Lett. {\bf 113} (2014) 252502. 

\bibitem{sieja} K. Sieja, Phys. Rev. Lett. {\bf 119} (2017) 052502.
\bibitem{sieja2} K. Sieja, EPJ Web of Conferences {\bf 146} (2017) 05004.

\bibitem{70ni} A. C. Larsen {\it et al.}, Phys. Rev. C {\bf 97} (2018) 054329. 

 \bibitem{midtbo} J. E. Midtb\o{}, A. C. Larsen, T. Renstr\o{}m, F. L. Bello Garrote, and E. Lima,  Phys. Rev. C {\bf 98} (2018) 064321.


\bibitem{GDRenergy}  J. J. Gaardh\o{}je, Annu. Rev. Nucl. Part. Sci. {\bf 42} (1992)  483.
\bibitem{GDRreview} K. A. Snover, Ann. Rev. Nucl. Part. Sci.  {\bf 36} (1986)  545.
\bibitem{schiller} A. Schiller and M. Thoennessen,  Atom. Data Nucl. Data Tables {\bf 93} (2007) 549.            

\bibitem{levinger3} J. S. Levinger and H. A. Bethe,  Phys. Rev. {\bf 78} (1950) 115.



\bibitem{hausser2} O. H\"ausser {\it et al.}, Nucl. Phys. A {\bf 212} (1973)  613.  
\bibitem{orce_10Be} J. N. Orce {\it et al.}, Phys. Rev. C {\bf 86} (2012)  041303(R).
\bibitem{12C_kumar} M. K. Raju {\it et al.}, Phys. Lett. B.  {\bf 777} (2018) 250. 

\bibitem{larsen_V} A. C. Larsen {\it et al.},  Phys. Rev. C {\bf 73} (2006) 064301. 




\bibitem{becvar} G. A. Bartholomew {\it et al.}, 
Adv. Nucl. Phys. {\bf 7} (1973) 229.

\bibitem{kheswa2} V. B. Kheswa, private communication. 

\bibitem{exfor} https://www-nds.iaea.org/exfor/exfor.htm
\bibitem{ensdf} https://www.nndc.bnl.gov/ensdf/								  
\bibitem{oslo} http://www.mn.uio.no/fysikk/english/research/\\about/infrastructure/OCL/nuclear-physics-research/compilation/



\bibitem{vessiere_Sc} A. Vessiere {\it et al.}, Nucl. Phys. A {\bf 227} (1974) 513. 
\bibitem{larsen_Sc} A. C. Larsen {\it et al.},  Phys. Rev. C {\bf 76}  (2007) 044303.
\bibitem{shoda} K. Shoda, Nucl. Phys. A {\bf 239} (1975) 397.
\bibitem{fultz_V} S. C. Fultz {\it et al.}, Phys. Rev {\bf 128} (1962) 2345. 

\bibitem{borodina_Fe}  S. S. Borodina {\it et al.}, Moscow State Univ. Inst. of Nucl. Phys. Rep.  (2000) 6.

\bibitem{Carlos_Ge} P. Carlos {\it et al.}, Nucl. Phys. A {\bf 258} (1976)  365.
\bibitem{spyrou_Ge} A. Spyrou {\it et al.}, Phys. Rev. Lett. {\bf 113} (2014) 232502.

\bibitem{Berman_Zr} B. L. Berman {\it et al.}, Phys. Rev. {\bf 162} (1967) 1098.
\bibitem{Guttormsen_Zr} M. Guttormsen {\it et al.}, Phys. Rev. C {\bf 96} (2017) 024313.
\bibitem{utshumiya_Zr} H. Utsunomiya {\it et al.}, Phys. Rev. Lett. {\bf 100} (2008) 162502. 

\bibitem{beil_Mo} H. Beil {\it et al.}, Nucl. Phys. A {\bf 227} (1974) 427.

\bibitem{utshumiya_Mo} H. Utsunomiya {\it et al.}, Phys. Rev. C {\bf 88} (2013) 015805.

\bibitem{beil_La} H. Beil {\it et al.},  Nucl. Phys. A {\bf 172} (1971)  426. 

\bibitem{carlos_152Sm}  P. Carlos {\it et al.}, Nucl. Phys. A {\bf 225} (1974)  171.







\bibitem{guttormsen2} M. Guttormsen {\it et al.}, Phys. Rev. C {\bf 68} (2003) 064306. 





\bibitem{lectures} R. Berg\`ere, Lecture Notes in Physics {\bf 61} (1977) 1.


\bibitem{orce_p} J. N. Orce, Phys. Rev. C {\bf 91} (2015) 064602.


\bibitem{bohigas2} O. Bohigas,  Lecture Notes in Physics {\bf 137} (1981) 65.


\bibitem{10B_vermeer} W. J. Vermeer {\it et al.}, Aust. J. Phys. {\bf 35} (1982)  283.
\bibitem{barker} F. C. Barker, Aust. J. Phys. {\bf 35} (1982)  291.











\bibitem{cebo} C. Ngwetsheni and J. N. Orce, in preparation (2019). 


\bibitem{balashov} V. V. Balashov, J. Exptl. Theoret. Phys. U.S.S.R. {\bf 42} (1962) 275. 












 















































%
%
%
%
%
%
%
%
%
%
%
%
%
%
%
%
%
%
%











%
%








%
%
%













\end{thebibliography}
\end{document}